\documentclass{article}

\usepackage{cite}
\usepackage[utf8]{inputenc}
\usepackage{xcolor, graphicx}
\usepackage{listings, lstcoq}

\usepackage{authblk}

\usepackage{pxfonts} 

\title{A Coq-based synthesis of Scala programs which are correct-by-construction}

\author[1,2]{Youssef El Bakouny}
\author[1]{Tristan Crolard}
\author[2]{Dani Mezher}

\affil[1]{CEDRIC - CNAM - Paris - France}
\affil[2]{CIMTI - ESIB - Saint-Joseph University - Beirut - Lebanon}

\date{May 1, 2017}

\begin{document}

\maketitle

\begin{abstract}
	The present paper introduces Scala-of-Coq, a new compiler that allows a Coq-based synthesis of Scala programs which are ``correct-by-construction''. A typical workflow features a user implementing a Coq functional program, proving this program's correctness with regards to its specification and making use of Scala-of-Coq to synthesize a Scala program that can seamlessly be integrated into an existing industrial Scala or Java application.
\end{abstract}

\lstset{
	frame=single,
	language=Scala,
	aboveskip=3mm,
	belowskip=3mm,
	showstringspaces=false,
	columns=flexible,
	basicstyle={\small\ttfamily},
	keywordstyle=\bfseries,
	numbers=none,
	breaklines=true,
	breakatwhitespace=true,
	tabsize=2
}

\section{Introduction}
In our modern world, software bugs are becoming increasingly detrimental to the engineering industry. Since software components are interconnected, a bug in one component is likely to affect others, causing a system-wide failure. Additionally, software bugs have often created system vulnerabilities that are exploitable by malicious users~\cite{DBLP:conf/oopsla/Pierce16}.

These bugs are essentially due to a lack of rigor in the software engineering field where the correctness of a program is thoroughly tested but rarely proven. In fact, current industrial software development practices rely solely on the expensive and time-consuming development of tests in an attempt to hunt down bugs. For non-trivial cases, this approach can never guarantee that a software is correct with regards to its specification.

As a result, we have recently witnessed interesting initiatives aimed at developing new science, technologies and tools capable of responding to the aforementioned software engineering challenges. A remarkable example of such an initiative is a U.S. National Science Foundation (NSF) expedition in computing project called ``the Science of Deep Specification (Deep Spec)''~\cite{DBLP:conf/oopsla/Pierce16}.

These initiatives make use of ``formal methods'', potentially as a complement to software testing, with the goal of providing the means of developing software that is correct-by-construction. These formal methods can be defined as a set of techniques and mathematical theories allowing both the specification of a software system in an unambiguous manner, and the rigorous verification of the conformity of this software system with regards to its specification. This verification can be done by proving that the program satisfies a set of lemmas or theorems constituting its specification.

\section{Proof assistants}
Since the manual checking of realistic program proofs is impractical or, to say the least, time-consuming; several proof assistants have been developed to provide machine-checked proofs. Isabelle/HOL~\cite{Nipkow2015} and Coq~\cite{Coq:manual} are currently the world's two leading proof assistants. 

Both of these proof assistants allow the interactive construction of formal proofs; Coq is based on a logical framework known as the calculus of inductive constructions~\cite{Pierce:SF}, while Isabelle/HOL is based on high order classical logic. Coq supports code generation in Objective Caml, Haskell and Scheme~\cite{Coq:manual}, while Isabelle/HOL supports code generation in Objective Caml, Haskell, SML and \emph{Scala}~\cite{Haftmann2010}. 

The code generation capabilities of these proof assistants enable the synthesis of programs which are correct-by-construction. However, they do not address the verification needs of existing software programs. This need is covered by other initiatives such as Leon~\cite{Blanc2013}, Why3~\cite{DBLP:conf/esop/FilliatreP13} and, also, Sireum Logika\footnote{http://logika.sireum.org}. Leon and Logika allow the automatic verification of \emph{Scala} programs. Leon can also resort to Isabelle/HOL machine-checked proofs when its automatic verification mechanism fails to give an answer. Similarly, Why3 allows the automatic verification of Objective Caml programs with the option of resorting to Coq on failure.

Coq has been successfully used to implement CompCert~\cite{DBLP:conf/cpp/Leroy12}, the world’s first and only formally verified C compiler. It is also widely used in several research projects including the aforementioned DeepSpec project, hence our interest in this particular proof assistant.

\section{The Scala Programming Language}
The Scala programming language~\cite{DBLP:journals/cacm/OderskyR14} was developed by Martin Odersky and his ``École Polytechnique Fédérale de Lausanne'' (EPFL) team in 2003. It is currently of high interest to several industrial projects thanks to its practical fusion of functional and object-oriented programming; as well as its seamless interoperability with Java.

Scala is the implementation language of many important frameworks, including Apache Spark, Kafka, and Akka. It also provides the core infrastructure for sites such as Twitter and Coursera.

Given the importance of Scala in the industrial world, the Coq proof assistant would highly benefit of a Scala code generation feature. For this purpose, the ``Scala-of-Coq'' compiler is currently being implemented.

\section{Compiling Coq to Scala}
The main objective of the Scala-of-Coq compiler is to allow the synthesis of Scala programs that are correct-by-construction. The compiler itself is written in Scala and its underlying Coq parser is largely based on the use of parser combinators~\cite{DBLP:journals/jfp/HuttonM98}, easing the construction of complex parsers out of primitive ones. The Scala library that supports these parser combinators~\cite{Moors2008} facilitated the implementation of the Coq parser without the need to resort to parser generators such as Lex/Yacc or ANTLR. This is allowing us to create and manipulate Coq’s Abstract Syntax Tree (AST) directly in Scala while taking full advantage of the language’s fusion of object-oriented and functional paradigms.

Using Scala-of-Coq, the Coq program in listing \ref{CoqProgram}, which computes the number of nodes in a binary tree, was successfully converted to the Scala program in listing \ref{ScalaCode}.

\begin{lstlisting}[language=Coq, caption={Source Coq Program}, captionpos=t, label=CoqProgram]
Inductive Tree : Type :=
  Leaf : Tree
| Node(l r : Tree): Tree.

Fixpoint size (t: Tree) : nat :=
match t with
  Leaf => 1
| Node l r => 1 + (size l) + (size r)
end.
\end{lstlisting}

\begin{lstlisting}[caption={Generated Scala Program}, captionpos=t, label=ScalaCode]
object CertifiedTree {
  sealed trait Tree
  case object Leaf extends Tree
  case class Node(l: Tree, r: Tree) extends Tree
  def size(t: Tree): BigInt =
    t match {
      case Leaf => 1
      case Node(l, r) => 1 + (size(l) + size(r))
    }
}
\end{lstlisting}

It is important to note that the source Coq program can be formally verified by proving the lemmas that define its specification. For example, the lemmas in listing \ref{CoqLemmas} were successfully proven using the definitions provided in listing \ref{CoqProgram}.

Once the source Coq program is formally proven with regards to its specification (portrayed as Lemmas), the generated Scala program can also be considered, assuming a few reasonable hypotheses, formally verified with regards to the same specification.

\begin{lstlisting}[language=Coq, caption={Coq Lemmas}, label=CoqLemmas]
Lemma size_left: forall l r : Tree, size (Node l r) > size l.
Lemma size_right: forall l r : Tree, size (Node l r) > size r.
\end{lstlisting}

\section{Acknowledgments}
The authors would like thank the National Council for Scientific Research in Lebanon (CNRS-L)\footnote{http://www.cnrs.edu.lb/} for their funding, as well as Murex S.A.S\footnote{https://www.murex.com/} for providing financial support.

\section{Conclusion}
In conclusion, the Scala-of-Coq compiler enables users of the Coq proof assistant to generate Scala programs that are formally verified. An example of a practical use case of Scala-of-Coq depicts a Coq user that generates a verified Scala component and seamlessly integrates it into an existing industrial Scala or Java application.

Future developments will focus on the verification of existing Scala programs using Coq. This verification should also include the support of imperative side effects and a restricted subset of parallel programs.

Finally, a framework for the support of additional languages, such as Swift, should be implemented.

\bibliography{coq-based-scala-program-extraction}{}
\bibliographystyle{plain}

\end{document}